\documentclass[preprint,showpacs,preprintnumbers,amsmath,amssymb, superscriptaddress,longbibliography,nofootinbib]{revtex4-1}

\usepackage{textcomp}
\usepackage{makeidx}
\usepackage{amsmath}
\usepackage{subfigure}
\usepackage{amssymb}
\usepackage{hyperref}
\usepackage{graphicx}
\usepackage{epstopdf}
\usepackage{color}
\usepackage{soul}

\begin{document}

\title{The Gravitational decoupling method: the higher dimensional case to find new analytic solutions}

\author{Milko Estrada}
\email{milko.estrada@gmail.com}
\affiliation{Department of Physics, Universidad de Antofagasta, 1240000 Antofagasta, Chile}

\author{Reginaldo Prado}
\email{rggixpf@gmail.com}
\affiliation{Department of Physics, Universidad de Antofagasta, 1240000 Antofagasta, Chile}

\date{\today}
\begin{abstract}
We provide a higher dimensional extension of the gravitational decoupling method. This extended method allows to obtain new analytic and well behaved solutions  that could be associated to higher dimensional stellar distributions. Furthermore, we find a new five dimensional anisotropic and well behaved analytical solution.  
\end{abstract}

\maketitle
\section{Introduction}
In recent years, great interest in astrophysical scenarios with anisotropic matter distribution has emerged. Ruderman \cite{Ruderman} says that, in realistic stellar models, the nuclear matter may have an anisotropic behavior in certain very high density ranges. Furthermore, Harko and Mak \cite{Harko1} say that, from the point of view of the Newtonian gravity, the spherical galaxies can have an anisotropic matter distribution.

These anisotropic models have been widely used in literature to describe compact objects ( for example in references \cite{Abreu,Gupta,Sunil,Newton,Harko1} ). For representing a compact object, the 
interior of a star have to comply with the general physical conditions listed bellow in section \ref{ejemplo} in which a well behaved anisotropic 
solution is described.

Recently, great interest in finding new analytic and anisotropic solutions for Einstein field equations has emerged. However, this is not a simple task due to the highly nonlinear behavior of these equations. 
Regarding this, Ovalle in (2017) \cite{Ovalle1} proposed a method called {\it Gravitational Decoupling of Sources} which applies a {\it Minimal Geometric Deformation} (MGD) to the temporal and radial metric components together with a {\it decoupling of sources}, and thus, leads to new anisotropic solutions. Since its appearance, it has been used to find several new anisotropic and well behaved solutions that represent stellar distributions \cite{Ovalle1,Ovalle2,Ovalle3,milko,Sharif1,Tello,Perez,Sharif2,Camilo,Tello1,Ovalle8,Tello3}. The last references mentioned are based on the application of the gravitational decoupling method to a well known and well behaved isotropic seed solution and, product of it, a new well behaved anisotropic solution is obtained. Besides, this method has been used to obtain new black hole solutions, as you can see in references \cite{Ovalle4,Contreras,Contreras2,Contreras4}, and other applications in references \cite{Contreras1,Angel,Ovalle9,Ovalle10,Contreras5,Contreras6}. Before of the gravitational decoupling method, the minimal geometric deformation of the space time was used to search solutions of the brane world models in references \cite{Ovalle5,Ovalle6,Ovalle7}.

In the gravitational decoupling method, the seed energy momentum tensor $\bar{T}_{\mu \nu}$ is deformed by an additional source $\theta_{\mu \nu}$,  whose coupling is proportional to the constant $\alpha$ and causes anisotropic effects on the self-gravitating system. This additional source can contain new fields, like scalar, vector and tensor fields \cite{Ovalle2}. Therefore the energy momentum tensor is:

\begin{equation}
    T_{\mu \nu} = \bar{T}_{\mu \nu} + \alpha \theta_{\mu \nu}, \label{EM}
\end{equation}
with the corresponding conservation equation:
\begin{equation}
    \nabla_\nu T^{\mu \nu} =0.
\end{equation}

The method is explained by Ovalle in reference \cite{Ovalle2}: ``given two gravitational sources: an isotropic source A and an extra source B, standard Einstein’s equations are first solved for A, and then a simpler set of {\it quasi-Einstein} equations are solved for B. Finally, the two solutions can be combined in order to derive the complete solution for the total system.".

Due to the above described ideas about the method, there are several open questions that arise in natural form:
\begin{itemize}
    \item Is it possible to use the gravitational decoupling method to obtain new solutions in scenarios with extra dimensions?.
    \item Is it possible to apply the gravitational decoupling method to a non isotropic ( {\it i.e.} anisotropic ) and non well behaved seed solution and, product of it, obtain a new well behaved and anisotropic solution (but different to from the seed) ?.
\end{itemize}

Although there are no experimental evidences of the presence of extra dimensions in our universe, there are some branches of theoretical physics that have studied these scenarios, for example, the string theory, branewold models and higher dimensional black holes (some examples in references \cite{Peru,Randall1,Randall2, milko2,Emparan,milko3}). However, there are not many studies in literature about higher dimensional compact objects because there are few higher dimensional physically viable solutions. 

The reference \cite{AltaEstrellaA} describes two well behaved and isotropic five dimensional solutions, where the first one corresponds to Einstein Gauss Bonnet theory and the second one corresponds to the Einstein Hilbert theory and represents a five dimensional generalization of the Durgapal-Bannerji model. In both solutions the isotropic criteria of physical acceptability are fulfilled, so the energy density and pressure are decreasing monotonically, whereas the pressure
vanishes at the boundary of the stellar distribution, and the sound velocity fulfills the causality. Other higher dimensional isotropic and well behaved solutions are found in reference \cite{AltaEstrellaB} for the Finch Skea model in Pure Gauss Bonnet theory, in reference \cite{AltaEstrellaC} for the Buchdahl-Vaidya Tikekar model in Pure Gauss Bonnet theory, and in reference \cite{AltaEstrellaD} for one ultra compact star in the Buchdahl-Vaidya Tikekar model in the Einstein Hilbert theory. 

The above mentioned examples describe isotropic and well behaved compact objects. However the reference \cite{AltaEstrellaE} (and references therein ) shows some theoretical arguments which indicate that the pressure inside a compact object is not essentially isotropic in nature.   

In the same reference \cite{AltaEstrellaE}, a well behaved and anisotropic higher dimensional solution of Einstein field equations in $5D$, $6D$ and $11D$ is shown by using a Lorentzian energy density of the Nazari and Mehdipour style. This solution fulfills the anisotropy criteria for physical admissibility listed bellow in subsection \ref{Lista}, and therefore could represent to a higher dimensional compact object. Other anisotropic and well behaved higher dimensional solution is seen in reference \cite{AltaEstrellaF} corresponding to the Find and Skea metric of a hydrodinamical stable star.

All the aforementioned examples were constructed using a generalization of a previously known model. In contrast, in this work we will apply the Gravitational Decoupled of Sources method to a non well behaved higher dimensional seed solution. If this higher dimensional seed solution is not physically well behaved , it could not have a physical interest by itself. Besides, it also could be a previously unknown solution. However, it is expected that the new anisotropic solution obtained from the application of the method to the seed solution be well behaved from the physical point of view, and thus it could represent a new toy model of higher dimensional anisotropic compact object. Furthermore, it is expected that the new solution be stable from Abreu \cite{Abreu} and adiabatic index criteria \cite{Herrera1,Heint1}.  

The work is organized as follows : Section \ref{seccion2} presents the Higher dimensional Einstein field equations for the energy momentum tensor \ref{EM}. In section \ref{metodo} we extend the Gravitational Decoupling of Sources method to the higher dimensional case and we define the matching conditions. Section \ref{ejemplo} is devoted to the application of the method to a non well behaved five dimensional solution and then, a new five dimensional and anisotropic well behaved solution that also fulfills with the criteria of stability \cite{Heint1,Herrera1,Abreu} is obtained. Finally, section \label{conclusion} summarizes the essential of this work and exposes some conclusions. 

\section{Higher dimensional Einstein equations for multiples sources} \label{seccion2}

The higher dimensional Einstein Hilbert equations are (to simplify the later graphic analysis we take $\kappa=1$):

\begin{equation}
    G^\mu_\nu = \kappa^2 T^\mu_\nu,
\end{equation}
where $T^\mu_\nu=\mbox{diag}(-\rho,p_r,p_\theta,p_\theta,...)$ is given by equation \ref{EM}, and where the seed energy momentum tensor is given by $\bar{T}^\mu_\nu= \mbox{diag}(-\bar{\rho},\bar{p}_r,\bar{p}_\theta,\bar{p}_\theta,...)$. Thus, it is easily see that:

\begin{align}
    \rho&=\bar{\rho}-\alpha \theta^0_0 \label{densidadefectiva} \\
    p_r &= \bar{p}_r +\alpha \theta^1_1 \label{presionradialefectiva} \\
    p_\theta &= \bar{p}_{\theta} +\alpha \theta^2_2 \label{presiontangencialefectiva}
\end{align}

So, since $\theta^1_1 \neq \theta^2_2 =\theta^3_3=...$ and $\bar{p}_r \neq \bar{p}_\theta=\bar{p}_\phi=...$ these sources induce an anisotropy:
\begin{equation}
    \Pi = \bar{p}_\theta-\bar{p}_r+ \alpha \Big ( \theta^2_2-\theta^1_1 \Big ) .\label{factoranisotropia}
\end{equation}

In Schwarzschild-like coordinates, a static $d$ dimensional spherically symmetric metric reads:
\begin{equation}
    ds^2=-e^{\nu(r)}+e^\lambda dr^2+r^2 d\Omega^2_{d-2}, \label{metrica1}
\end{equation}
where $r$ run from $r=0$ (center of the object) to $r=R$ (surface of the object).

So, the $(t,t)$ and $(r,r)$ components of the equation of motions are given by:

\begin{equation}
    \bar{\rho}-\alpha \theta^0_0 = \frac{d-2}{2} \frac{\Big(r^{d-3}(1-e^{-\lambda}) \Big )'}{r^{d-2}} \label{tt}
\end{equation}

\begin{equation}
    \bar{p}_r + \alpha \theta^1_1   = (d-2) \frac{\Big ( e^{-\lambda} \big(r \nu(r)'+(d-3) \big)-(d-3) \big ) \Big )}{2 r^2}.  \label{rr}
\end{equation}

For an energy momentum tensor of the form $T^\mu_\nu = \mbox{diag} (-{\rho}, p_r, p_\theta, p_\phi, ...)$. From spherical symmetry we have for all the $(d-2)$ angular coordinates $p_t=p_\theta=p_\phi=...$ and, the conservation law $T^{AB}_{;B}=0$ gives:

\begin{equation}
\frac{1}{2} (p_r+ {\rho}) \nu'+p'_r+\frac{d-2}{r}(p_r-p_t ) =0 . \label{conservacion1}
\end{equation}

We solve the $(t,t)$ and $(r,r)$ components of the Einstein Field equations together with the conservation equation. Using  the Bianchi identities, we ignore the remaining $(\theta,\theta)=(\phi,\phi)=...$ components (the suspense points indicate that all the tangential components of the Einstein equations are similar). 

Inserting equations \ref{densidadefectiva}, \ref{presionradialefectiva} and \ref{presiontangencialefectiva} into the equation \ref{conservacion1} :

\begin{eqnarray} \label{conservacion11}
\frac{1}{2} (\bar{p}_r+\bar{\rho}) \nu'+\bar{p}_r'+ \frac{d-2}{r}(\bar{p}_r-\bar{p}_t ) + \alpha \Big ( \frac{1}{2} (\theta^1_1 - \theta^0_0) \nu'+(\theta^1_1)'+\frac{d-2}{r}(\theta^1_1-\theta^2_2 ) \Big )=0 \label{conservacion2}
\end{eqnarray}

\section{ Gravitational decoupling by MGD in the higher dimensional case} \label{metodo}

This method was initially proposed in references \cite{Ovalle1,Ovalle2} for the four dimensional case. Starting with a solution to equations \ref{tt}, \ref{rr} and \ref{conservacion2} with $\alpha=0$, namely {\it seed solution} $\{\eta,\mu,\bar{\rho},\bar{p}_r,\bar{p}_t \}$, where $\eta$ and $\mu$ are the corresponding metric functions:

\begin{equation}
    ds^2=-e^{\eta(r)}dt^2+\mu(r)^{-1} dr^2+r^2 d\Omega^2_{d-2}. \label{metrica2}
\end{equation}

Turning on the parameter $\alpha$, the effects of the source $\theta_{\mu \nu}$ appear on the seed solution $\{\eta,\mu,\bar{\rho},\bar{p}_r,\bar{p}_t \}$. These
effects can be encoded in the geometric deformation undergone by the seed fluid geometry $\{\eta,\mu \}$ in equation \ref{metrica2} as follows:
\begin{equation}
    \eta(r) \to \nu(r)=\eta(r) \label{deformaciontemporal}
\end{equation}

\begin{equation}
    \mu (r) \to e^{-\lambda} = \mu (r) + \alpha g(r). \label{deformacionradial}
\end{equation}

It means that only the radial component of the line element \ref{metrica2} is deformed, where $g(r)$ is the corresponding deformation of the radial part. Thus, replacing equations \ref{deformaciontemporal} and \ref{deformacionradial} into of equations \ref{tt}, \ref{rr} and \ref{conservacion2} , the system
splits into two sets of equations: 

\begin{enumerate}
    \item The standard higher dimensional Einstein equations for a seed solution (with $\alpha=0$), where $\eta(r) = \nu(r)$:
    
    \begin{equation}
    \bar{\rho}= \frac{d-2}{2} \frac{\Big(r^{d-3}(1-\mu) \Big )'}{r^{d-2}} \label{tt1}
\end{equation}

\begin{equation}
    \bar{p}_r  = (d-2) \frac{\Big ( \mu(r) \big(r \nu(r)'+(d-3) \big)-(d-3) \big ) \Big )}{2 r^2}.  \label{rr1}
\end{equation}
and the respective conservation equation:    
    \begin{eqnarray}
\frac{1}{2} (\bar{p}_r+\bar{\rho}) \nu'+\bar{p}_r'+ \frac{d-2}{r}(\bar{p}_r-\bar{p}_t )=0 \label{conservacion3}
\end{eqnarray}

\item The terms of order $\alpha$ give rise to the following { \it quasi-Einstein equations} \cite{Ovalle2}, which include the source $\theta_{\mu \nu}$:
     \begin{equation}
    \theta^0_0= \frac{d-2}{2} \frac{\Big(r^{d-3} g \Big )'}{r^{d-2}} \label{tt2}
\end{equation}

\begin{equation}
    \theta^1_1  = (d-2) \frac{\Big ( g \big(r \nu(r)'+(d-3)  \big ) \Big )}{2 r^2}.  \label{rr2}
\end{equation}

Inserting equation \ref{conservacion3} into equation \ref{conservacion11}, we can see the respective conservation equation:
 \begin{eqnarray}
  \frac{1}{2} (\theta^1_1 - \theta^0_0) \nu'+(\theta^1_1)'+\frac{d-2}{r}(\theta^1_1-\theta^2_2 ) =0 \label{conservacion4}
\end{eqnarray}
\end{enumerate}

It is worth stressing that equations \ref{conservacion3} and \ref{conservacion4} imply that there is no exchange of energy momentum between the fluid of seed solution and the extra source $\theta_{\mu \nu}$. So there is only purely gravitational interaction.

\subsection{Matching Conditions}
This subsection presents the matching conditions for the coupling of our new anisotropic solution (obtained from the Gravitational Decoupling method) with the higher dimensional Schwarzschild solution: 

\begin{equation}
    ds^2=-f(r)dt^2+f(r)^{-1}dr^2+r^2 d\Omega^2_{d-2},
\end{equation}
where 
\begin{equation}
    f(r)=1- \frac{M}{r^{d-3}}.
\end{equation}

Using the Israel-Darmois matching conditions, the first fundamental form says that there must be no jumps in the metric, therefore:

\begin{equation}
    \left[ds^{2}\right]_{\Sigma}=0, \label{firstform}
\end{equation}
where $\left[F\right]_{\Sigma}\equiv F\left(r \rightarrow R^{+}\right)- F\left(r \rightarrow R^{-}\right)$ represents the jump of any function $F=F(r)$ at the stellar surface $\Sigma$ (defined by $r=R$).

\begin{equation}\label{firstnu}
    g_{tt}^{\,\,-}(R)=g_{tt}^{\,\,+}(R),
\end{equation}
that yields to:
\begin{equation}
    e^{\nu(R)} = 1- \frac{M}{R^{d-3}} \label{matchingtemporal}
\end{equation}
and:
\begin{equation}\label{firstlambda}
    g_{rr}^{\,\,-}(R)=g_{rr}^{\,\,+}(R),
\end{equation}
that yields to:
\begin{equation}
    \mu(R)+\alpha g(R) = 1- \frac{M}{R^{d-3}} \label{matchingradial}
\end{equation}

The second fundamental form of Israel-Darmois, at the stellar surface (where $r=R$) gives \cite{Israel}:
\begin{equation}\label{secondform}
    \left[{G}_{\mu\nu}r^{\nu}\right]_{\Sigma}=0, 
\end{equation}
where $r_{\nu}$ is an unit radial vector. Using equation (\ref{secondform}) and the general Einstein equations, we find that:
\begin{equation}
    \left[{T}_{\mu\nu}r^{\nu}\right]_{\Sigma}=0,
\end{equation}
which, taking into consideration that the exterior geometry is vacuum, leads to:
\begin{equation}
    \bar{p}_r(R)+\alpha \theta^1_1(R)=0. \label{matchingpresion}
\end{equation}

\section{A very simple example of a new analytical and anisotropic well behaved higher dimensional solution.} \label{ejemplo}
In this section we apply the higher dimensional gravitational decoupling method to a non well behaved and non isotropic (anisotropic) seed solution in the five dimensional case $4+1$. In our case, the seed solution has no physical interest by itself. However, later on we will show that the new anisotropic solution obtained from the application of our method is well behaved from the physical point of view, and different from the seed solution.

\subsection{Seed solution}

Our seed solution corresponds to the equations \ref{tt1}, \ref{rr1} and \ref{conservacion3}, where $\alpha=0$:

\begin{align}
    \nu(r)=& \frac{1}{2}Cr^2+D     \\
    \mu(r)=&1-\Omega r^2      \\
    \bar{\rho}=& \frac{(d-1)(d-2)}{2}\Omega   \\
    \bar{p}_r=& \frac{d-2}{2} \Big ( C-(d-3) \Omega \Big ) - \frac{d-2}{2} \Omega C r^2   \\
    \bar{p}_\theta=& \frac{1}{4}C^2 r^2 - \frac{d-1}{2}C \Omega r^2 - \frac{1}{4}\Omega C^2 r^4 + \frac{d-2}{2}C-\frac{(d-2)(d-3)}{2}\Omega ,
\end{align}
where $d=5$ and $C,D$ and $\Omega$ are constants. This solution is not well behaved, since the energy density $\bar{\rho}$ has a constant value and, thus, is not monotonically decreasing. The energy density has the form of the $d$ dimensional cosmological constant, however, in our case the constant $\Omega$ represents only a constant value and, therefore, there is not a cosmological constant.

\subsection{Applying MGD to seed solution}
In the references \cite{Ovalle1,Ovalle2,Ovalle3,milko,Sharif1,Tello,Perez,Sharif2,Camilo}, mimic constraints were chosen for the components $\theta^0_0$ and $\theta^1_1$. In our work we impose a mimic constraint for the radial deformation $g(r)$. In the denominator of equations \ref{tt2} and \ref{rr2}, there are terms of the form $r^2$, therefore a suited election is such that these singularities are avoided. So, a simple mimic constraint is:

\begin{equation}
    g(r)=Ar^4+Br^2, \label{mimic}
\end{equation}
so, from equations \ref{deformaciontemporal} and \ref{deformacionradial}, the temporal and radial components of the metric are:

\begin{equation}
    \nu(r)=\frac{1}{2}Cr^2+D, \label{metricatemporal}
\end{equation}
and
\begin{equation}
    e^{-\lambda}=1-\Omega r^2 + \alpha \Big (  Ar^4+Br^2  \Big ) \label{metricaradial}
\end{equation}

Imposing the mimic constraint \ref{mimic} into equations \ref{tt2}, \ref{rr2} and \ref{conservacion4} for the five dimensional case $d=5$:
\begin{align}
    \theta^0_0=& 9Ar^2+6B \\
    \theta^1_1=& \frac{3}{2} (Ar^2+B)(Cr^2+2) \\
    \theta^2_2=& \frac{1}{4}AC^2r^6+\frac{1}{4}C(BC+10A)r^4+(2BC+5A)r^2+3B .
\end{align}

So, from equations \ref{densidadefectiva} ,\ref{presionradialefectiva} and \ref{presiontangencialefectiva}, the effective energy density, effective radial pressure and effective tangential pressure are, respectively:
\begin{align}
    \rho=& 6\Omega - \alpha \Big ( 9Ar^2+6B \Big) \\
    p_r=& \frac{3}{2}C-3\Omega-\frac{3}{2}C\Omega r^2 +\alpha \Big ( \frac{3}{2} (Ar^2+B)(Cr^2+2)  \Big ) \\
    p_\theta=& \frac{1}{4}C(C-8\Omega)r^2-\frac{1}{4}C^2\Omega r^4 + \frac{3}{2}C-3\Omega + \alpha \Big ( \frac{1}{4}C^2 A r^6 + \frac{1}{4}C(BC+10A)r^4 \nonumber \\
             &+(2BC+5A)r^2+3B \Big )
\end{align}

\subsection{Matching conditions}

Our matching conditions are given by equations \ref{matchingtemporal}, \ref{matchingradial} and \ref{matchingpresion}. The condition \ref{matchingtemporal} leads straightly to the value of the constant $D$. The conditions \ref{matchingradial} and \ref{matchingpresion} yield to:

\begin{equation}
    1-\Omega R^2 + \alpha \Big (  AR^4+BR^2  \Big ) = 1- \frac{M}{R^{d-3}},  \label{matchingA}
\end{equation}
and
\begin{equation}
    p_r(r=R)= \frac{3}{2}C-3\Omega-\frac{3}{2}C\Omega R^2 +\alpha \Big ( \frac{3}{2} (AR^2+B)(CR^2+2) \Big ) =0. \label{matchingB}
\end{equation}
 
Taking as free parameters the constants $\alpha$, $M$ and $R$, there are two equations and four unknown quantities $A$,$B$,$C$ and $\Omega$. So, in arbitrary way, in our example, we will impose two additional conditions:

\begin{itemize}
    \item The anisotropy factor \ref{factoranisotropia} has a positive and arbitrary value $N_1$ at the stellar surface $r=R$. This constraint has  sense, since the anisotropy factor is zero in the origin. Therefore, for obtaining a physically acceptable model,  one could think that this factor increases as $r$ approaches to the surface.
    \begin{equation}
        p_t(r=R)-p_r(r=R)=p_t(r=R)=N_1. \label{arbitrario1}
    \end{equation}
    \item The value of the square of tangential speed of sound has a positive and arbitrary value $N_2<1$ at the stellar surface $r=R$. This is consistent with the causality. 
    \begin{equation}
        v_t^2(r=R)= \frac{dp_t}{d\rho}(r=R)=N_2 \label{arbitrario2}
    \end{equation}
\end{itemize}

\subsection{Physical analysis of our example} \label{Lista}
We use in arbitrary way the values of mass and radius of the compact star RXJ 1856-37 $M=0,9041$ solar mass and $R=6$ KM \cite{Sunil}. In all the graphics the used constants $A$,$B$,$C$ and $\Omega$ are solutions of the  system \ref{matchingA}, \ref{matchingB}, \ref{arbitrario1} and \ref{arbitrario2}. 

The physical features of our example, which correspond to a well behaved compact distribution, are the following :

\begin{enumerate}
\item The metric components are regular inside the stellar distribution. In our case from equation \ref{metricatemporal} $e^\nu (r=0) \neq 0$ and from equation \ref{metricaradial} $e^{-\lambda}(r=0)=1$. Furthermore both components are free of singularities in all the range $r \in [0,R]$
    \item The energy density, and the radial and tangential pressures are positive and monotonic decreasing functions of $r$. Furthermore, the radial pressure vanishes at the boundary $r=R$. In figure \ref{presiones} these requirements are fulfilled. 
    
    \begin{figure}
\centering
\subfigure[Radial pressure .]{\includegraphics[width=85mm]{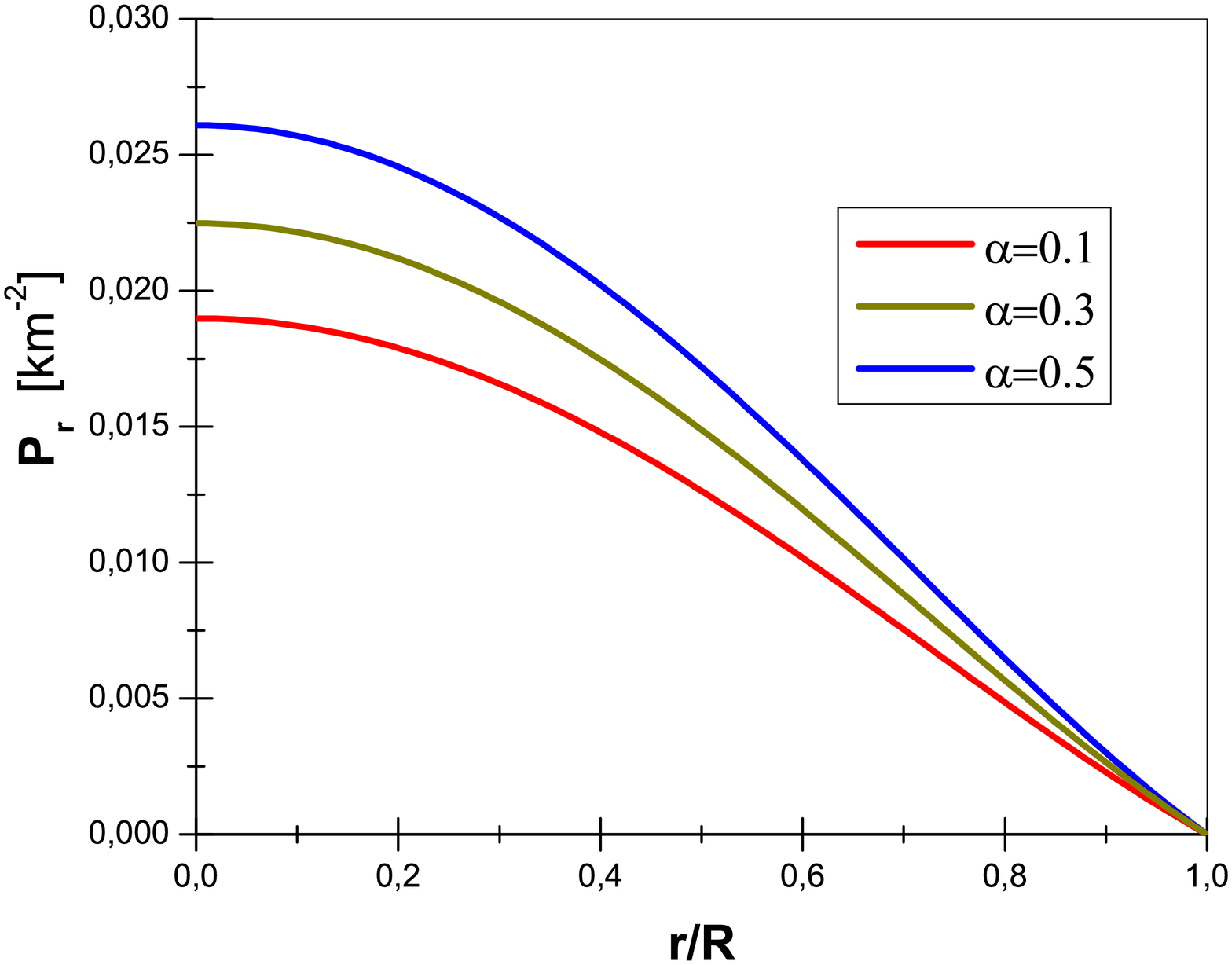}} 
\subfigure[Tangential pressure.]{\includegraphics[width=85mm]{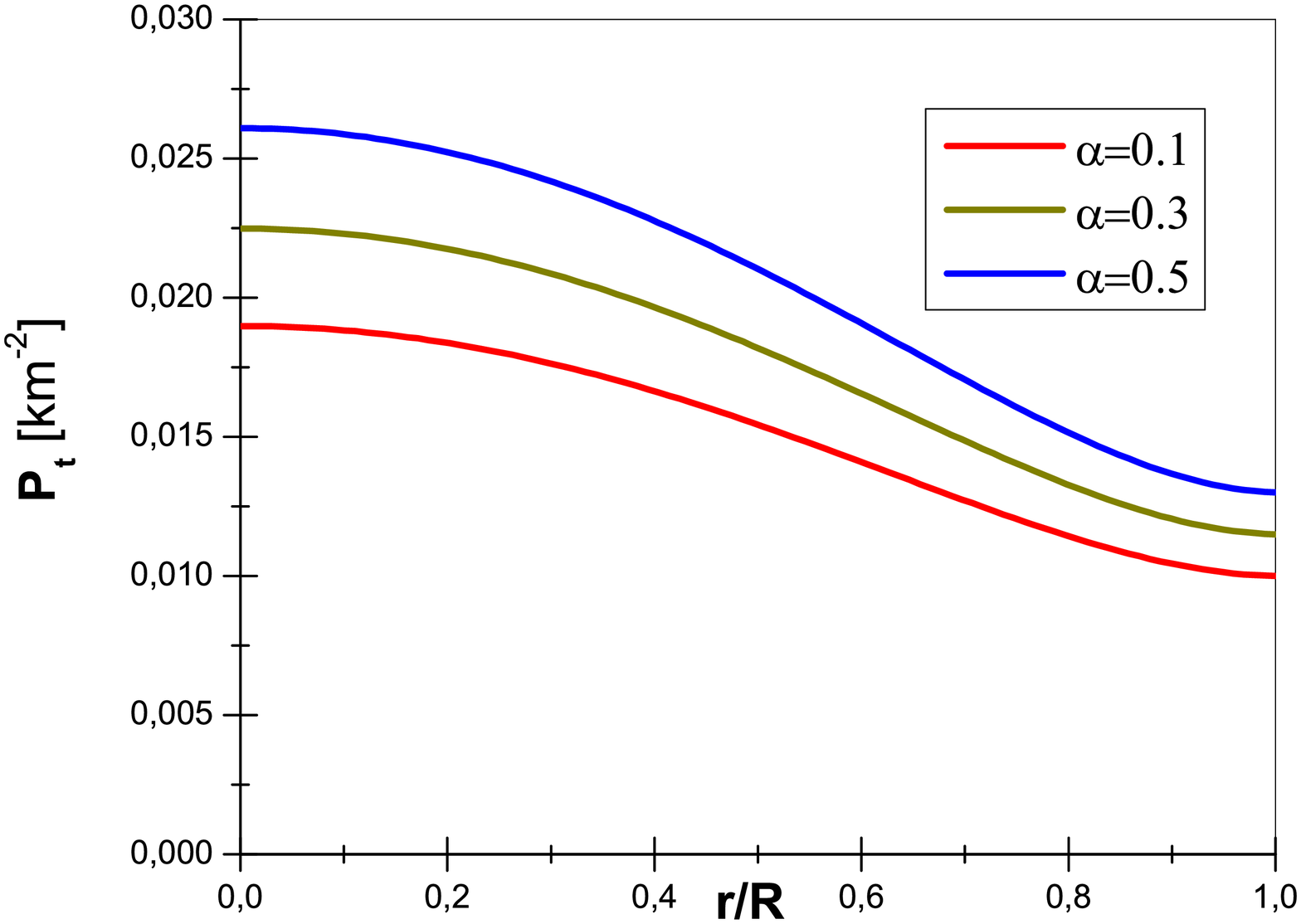}} 
\subfigure[Energy Density.]{\includegraphics[width=85mm]{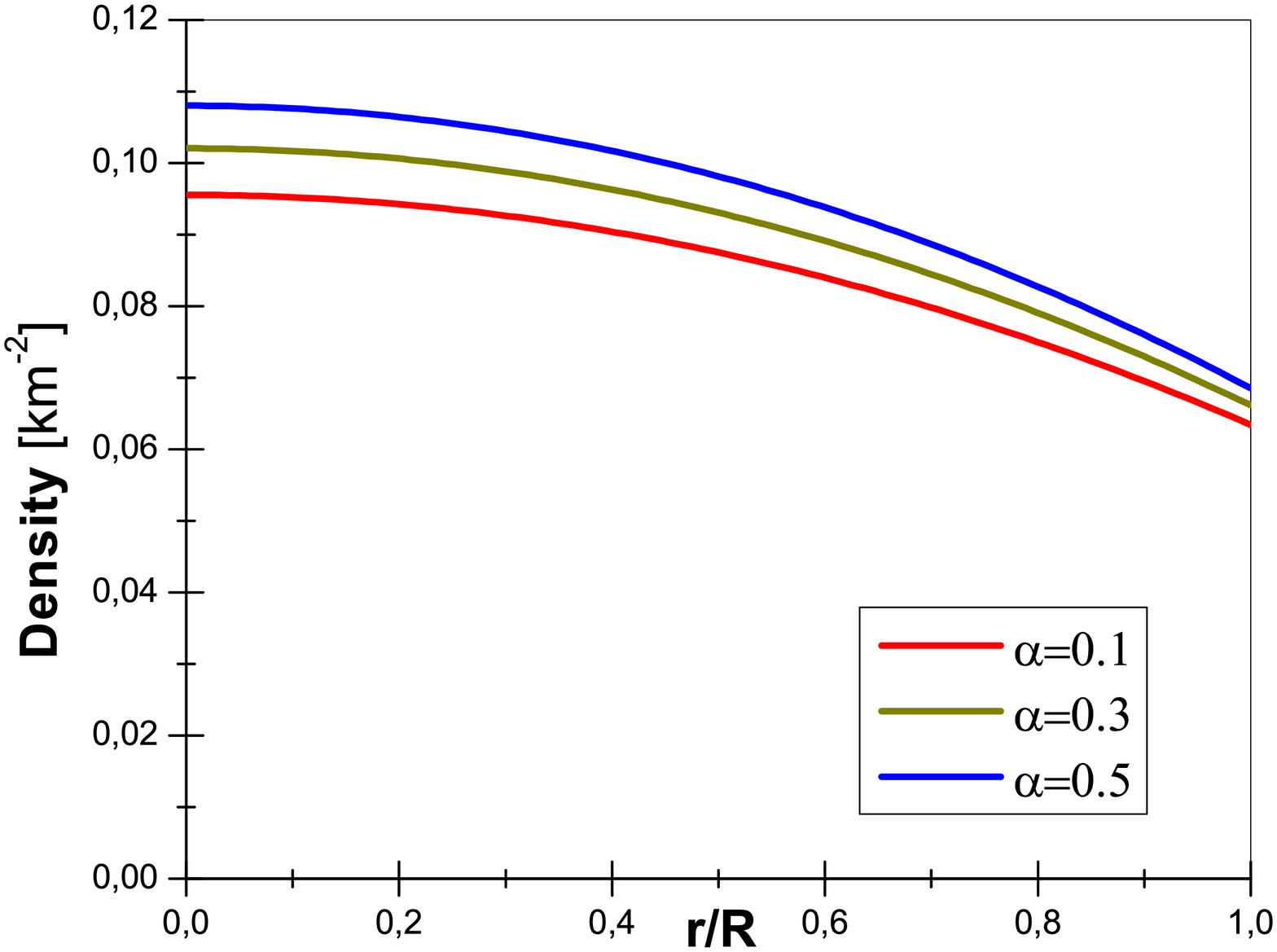}} 
\caption{$p_r$, $p_t$ and $\rho$ for $\alpha=0,1$ (with $N_1=N_2=10^{-2}$), $\alpha=0,3$ (with $N_1=N_2=1,15 \cdot 10^{-2}$) and $\alpha=0,5$ (with $N_1=N_2=1,3 \cdot 10^{-2}$) .}
\label{presiones}
\end{figure}
    
    \item The anisotropy factor vanishes at the origin and increase as $r$ also increase. For $r>0$ the tangential pressure is greater than radial pressure $p_t>p_r$. The behavior of the anisotropy factor is displayed in figure \ref{figanisotropia}.
    
\begin{figure}
\centering
{\includegraphics[width=85mm]{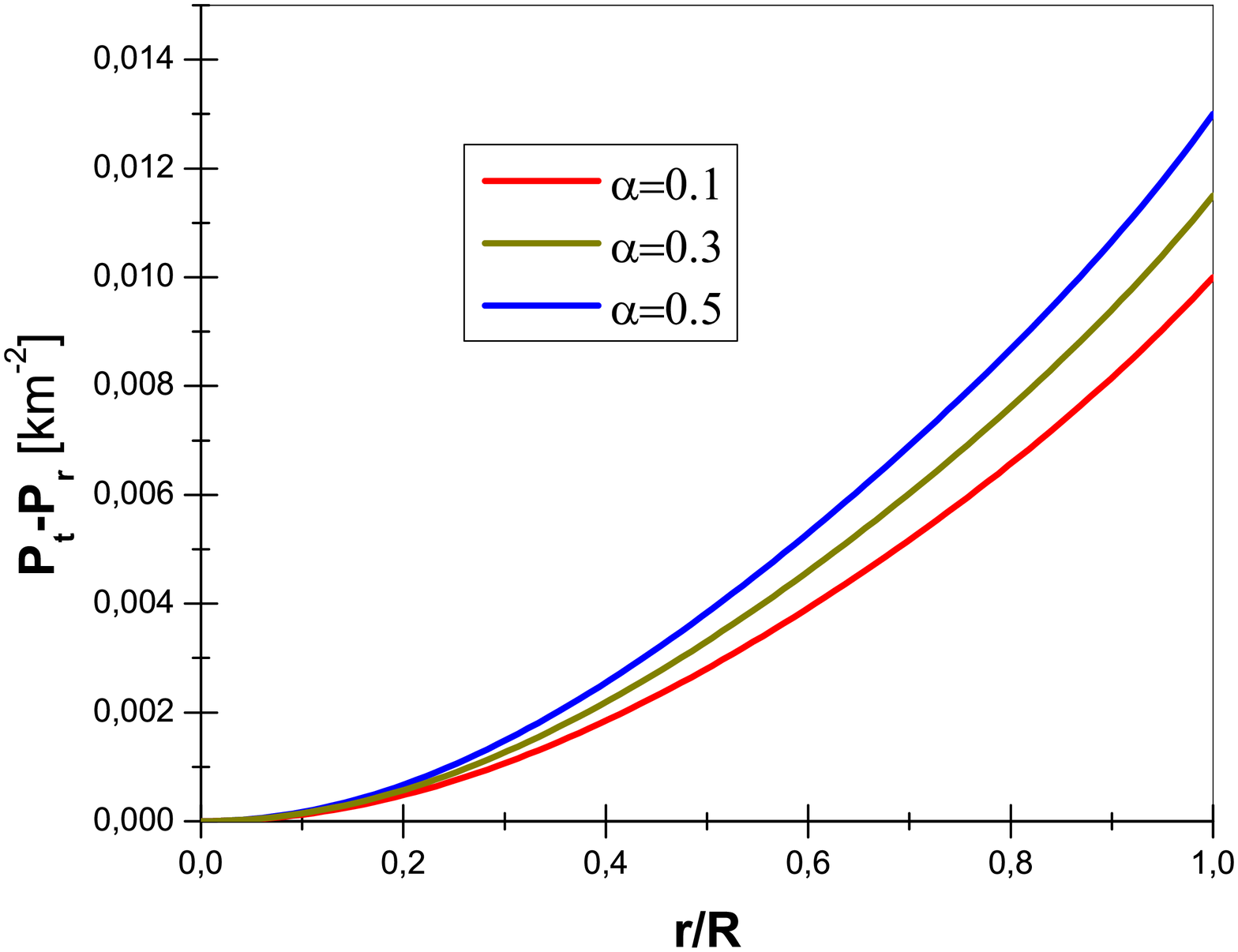}} 
\caption{Anisotropy factor for $\alpha=0,1$ (with $N_1=N_2=10^{-2}$), $\alpha=0,3$ (with $N_1=N_2=1,15 \cdot 10^{-2}$) and $\alpha=0,5$ (with $N_1=N_2=1,3 \cdot 10^{-2}$) .}
\label{figanisotropia}
\end{figure}
    
    \item The energy conditions: null energy condition (NEC), weak energy condition (WEC), dominant energy condition (DEC) and strong energy condition (SEC) are fulfilled:
    \begin{align}
        \mbox{NEC}&:\rho (r) \ge 0 , \\
        \mbox{WEC}&: \rho(r)-p_r(r) \ge 0 \mbox{   and  } \rho(r)-p_t(r) \ge 0, \\
        \mbox{DEC}&: \rho(r) \ge |p_r(r)|,|p_t(r)|, \\
        \mbox{SEC}&: \rho(r)-p_r(r)-2p_t(r) \ge 0 .
    \end{align}
    
    We see by simple inspection that NEC, WEC and DEC are fulfilled in figure \ref{presiones}, and SEC in figure \ref{figsec}. 
    
    \begin{figure}
\centering
{\includegraphics[width=85mm]{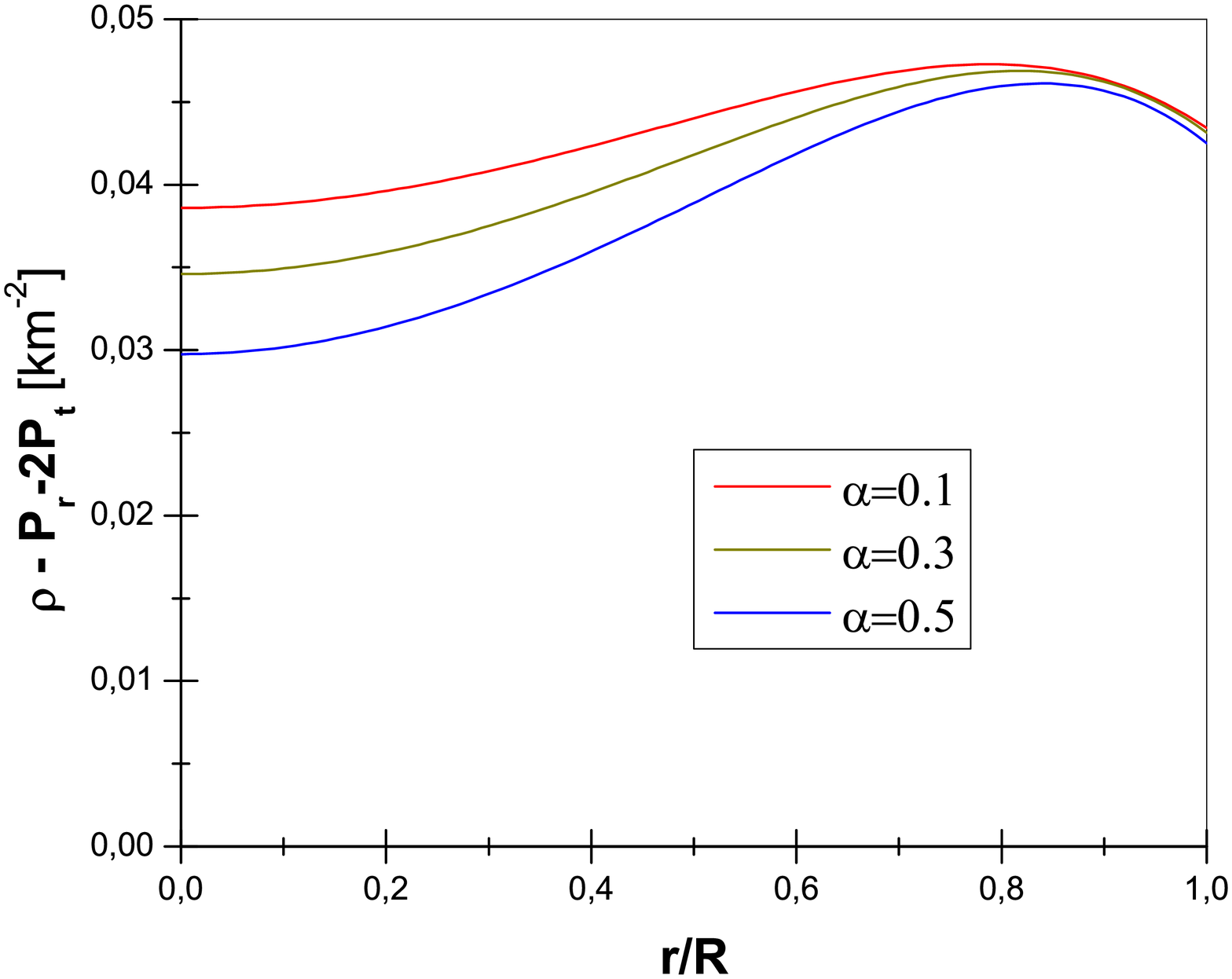}} 
\caption{SEC for $\alpha=0,1$ (with $N_1=N_2=10^{-2}$), $\alpha=0,3$ (with $N_1=N_2=1,15 \cdot 10^{-2}$) and $\alpha=0,5$ (with $N_1=N_2=1,3 \cdot 10^{-2}$) .}
\label{figsec}
\end{figure}
    
    \item The causality condition says that the radial and tangential components of the speed of sound are less than one. Therefore:
    \begin{align}
        v_r^2=& \frac{dp_r}{d\rho} <1 \\
        v_t^2=& \frac{dp_t}{d\rho} <1 .
    \end{align}
    
   In figure \ref{figvelocidad}, both components of the velocity are monotonically decreasing functions of $r$ and less than one.
    
     \begin{figure}
\centering
\subfigure[ Square of radial velocity .]{\includegraphics[width=85mm]{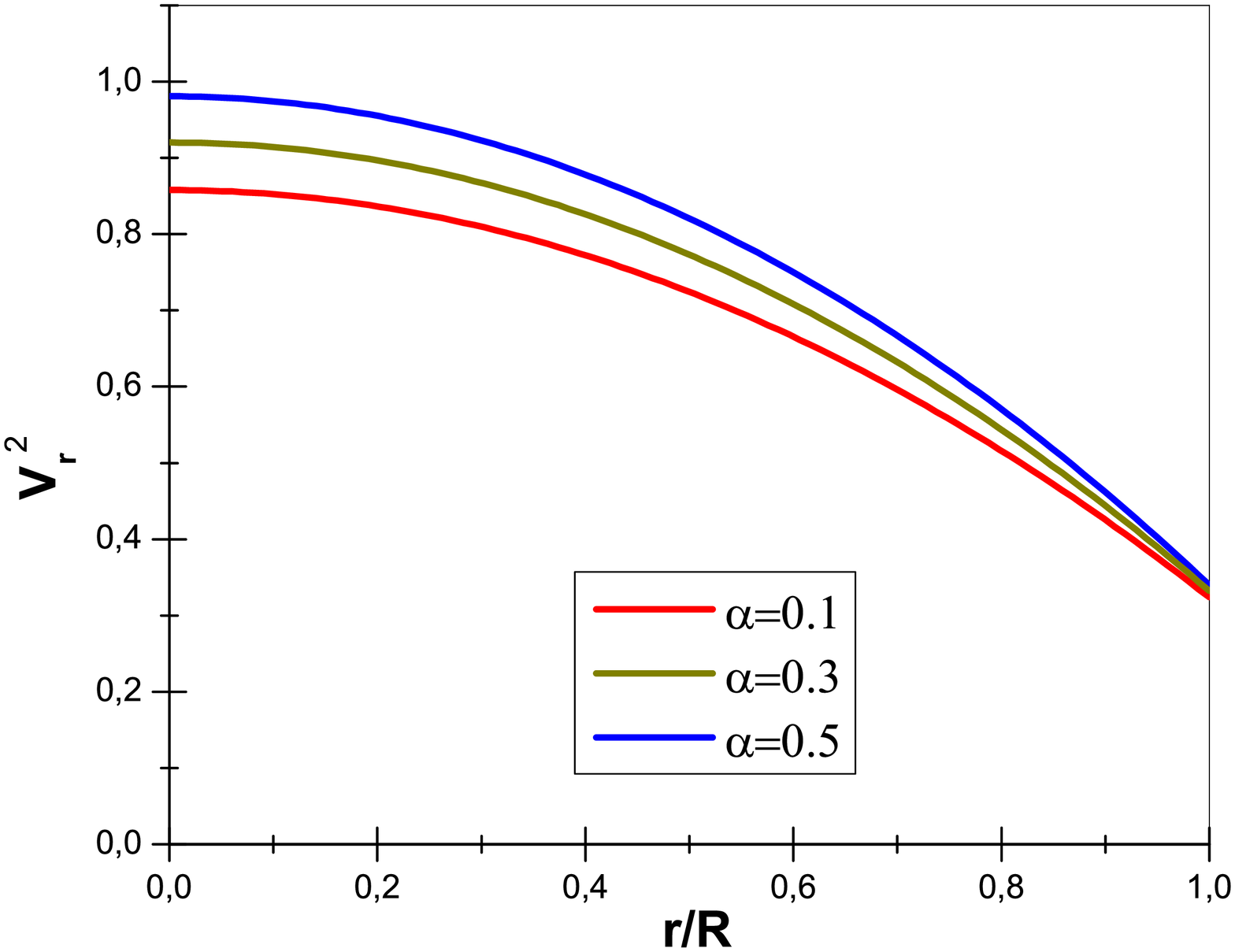}} 
\subfigure[ Square of tangential pressure.]{\includegraphics[width=85mm]{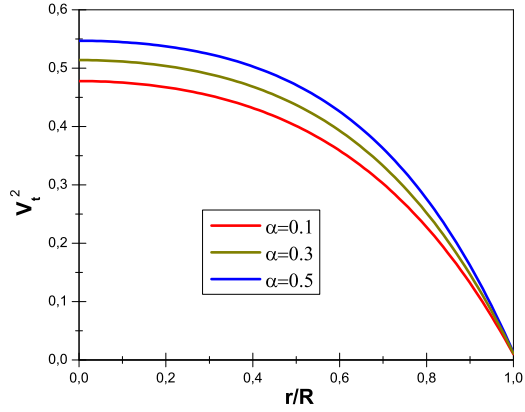}} 
\caption{$v_r^2$ and $v_t^2$ for $\alpha=0,1$ (with $N_1=N_2=10^{-2}$), $\alpha=0,3$ (with $N_1=N_2=1,15 \cdot 10^{-2}$) and $\alpha=0,5$ (with $N_1=N_2=1,3 \cdot 10^{-2}$) .}
\label{figvelocidad}
\end{figure}
    
\item Stability condition: We analyze the stability of our example through the two following criteria: 
\begin{enumerate}
    \item Abreu criterion: This criterion was development by Abreu et.al in reference \cite{Abreu}. This says that the model is potentially stable when $0<v_r^2-v_t^2<1$. In figure \ref{figabreu} this condition is fulfilled, therefore our model is stable. 
    
\begin{figure}
\centering
{\includegraphics[width=85mm]{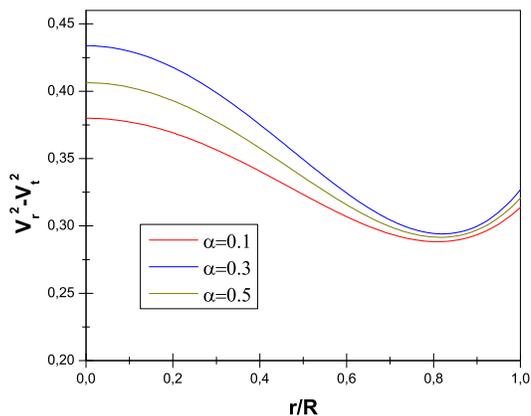}} 
\caption{$v_r^2-v_t^2$ for $\alpha=0,1$ (with $N_1=N_2=10^{-2}$), $\alpha=0,3$ (with $N_1=N_2=1,15 \cdot 10^{-2}$) and $\alpha=0,5$ (with $N_1=N_2=1,3 \cdot 10^{-2}$) .}
\label{figabreu}
\end{figure}
    
    \item Adiabatic index: The stability of an anisotropic spherical distribution also can be described by the adiabatic index, defined as \cite{Herrera1}:
    \begin{equation}
        \Gamma_r=\frac{\rho+p_r}{p_r}\frac{dp_r}{d\rho},
    \end{equation}
regarding this, Heintzmann and Hillebrandt \cite{Heint1} showed that the condition for a stable compact object is given by $\Gamma_r>4/3$. In figure \ref{figadiabatico} this condition is fulfilled and therefore our solution is stable.

\begin{figure}
\centering
{\includegraphics[width=85mm]{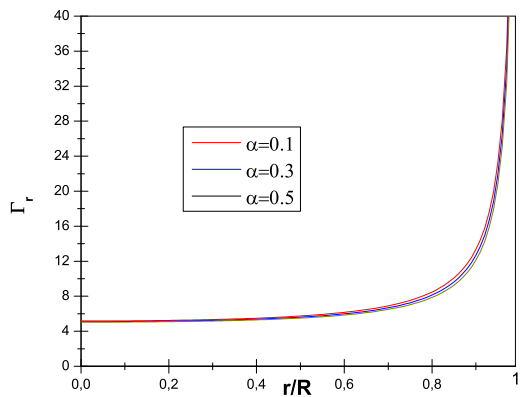}} 
\caption{$\Gamma_r$ for $\alpha=0,1$ (with $N_1=N_2=10^{-2}$), $\alpha=0,3$ (with $N_1=N_2=1,15 \cdot 10^{-2}$) and $\alpha=0,5$ (with $N_1=N_2=1,3 \cdot 10^{-2}$) .}
\label{figadiabatico}
\end{figure}

\end{enumerate}

\end{enumerate}

\section{Conclusions and remarks} \label{conclusion}
We have shown that it is possible to extend the gravitational decoupling method to a higher dimensional scenario. Beside, it is possible to apply this method to an anisotropic and non well behaved seed solution and, product of this, to obtain other different anisotropic solution, but well behaved from the physical point of view. 

In the same way as the four dimensional case, the obtained final solution is the result of the decoupling of Einstein equations in a seed sector described by the energy momentum tensor $\bar{T}^\mu_\nu$ and the quasi Einstein equations described by the source $\Theta^\mu_\nu$. So, the seed and the extra sources are separately conserved. Therefore, the combination of these two sectors only has gravitational interaction and does not have exchange of energy momentum \cite{milko}.

Applying the higher dimensional gravitational decoupling method to an anisotropic and non well behaved seed solution $\{\nu,\mu, \bar{\rho},\bar{p}_r, \bar{p}_t\}$, we have obtained a new five dimensional and well behaved anisotropic solution $\{\nu,\mu+\alpha g, {\rho},{p}_r, {p}_t$\}. This same example in dimensions greater than five could be studied in elsewhere. Unlike other higher dimensional and well behaved solutions previously studied in literature, our solution is not a generalization of other previously known model. Our seed solution is unknown and has no physical interest. However, applying the method to this seed solution, we have obtained a new  well behaved solution. We have used typical mass and radius values of compact objects corresponding to the star RXJ 1856-37 (for different values of $\alpha$). This choice is merely arbitrary because there is no evidence that the mentioned compact star be fully described by our solution in the higher dimensional case. Our solution is well behaved because it fulfills the following criteria of physical admissibility \cite{Newton}, and therefore could be a new toy model of anisotropic compact stars: 

\begin{itemize}
\item The metric components are free of physical singularities inside the stellar distribution.
\item The density and pressures are positive inside the star. Furthermore, these variables are decreasing functions.
\item The radial pressure drops from its maximum value (at $r=0$) to zero at the boundary ($r=R$).
\item  The solution satisfies the NEC, WEC, DEC and SEC conditions.                                
\item Inside the stellar distribution, the radial and tangential components of the speed of sound are less than the speed of light. Therefore the sound propagation is causal.
\end{itemize}

Additionally, our example shows a stable solution by using Abreu \cite{Abreu} and adiabatic index criteria \cite{Herrera1,Heint1}.

Despite the fact that in our example we have obtained a well behaved anisotropic solution, our higher dimensional algorithms, shown in section \ref{metodo}, could serve for other applications. Motivated by the fact that this method in four dimensions has also been used to find new black hole solutions \cite{Ovalle4,Contreras,Contreras2,Contreras4}, perhaps our higher dimensional method could serve for finding new higher dimensional black hole solutions in elsewhere.

\bibliography{mybib}

\end{document}